\documentclass[twoside]{article}
\usepackage{qic,epsfig}
\newcommand{\ket}[1]{\mbox{$ | #1 \rangle $}}

\begin{document}
\setlength{\textheight}{8.0truein}    

\runninghead{Quantum direct communication with mutual authentication} 
            {C.-A. Yen, S.-J. Horng, H.-S. Goan, T.-W.
            Kao, and Y.-H. Chou} 

\normalsize\textlineskip
\thispagestyle{empty}
\setcounter{page}{376}

\copyrightheading{9}{5\&6}{2009}{0376--0394}

\vspace*{0.88truein}

\alphfootnote

\fpage{376}

\centerline{\bf
QUANTUM DIRECT COMMUNICATION WITH MUTUAL AUTHENTICATION}
\vspace*{0.37truein}

\centerline{\footnotesize
Cheng-An Yen$^{1,}$\footnote{Electronic
  address: D9215006@mail.ntust.edu.tw} ,
Shi-Jinn Horng$^{1,2}$,
Hsi-Sheng Goan$^{3,}$\footnote{Electronic
  address: goan@phys.ntu.edu.tw (corresponding author)} ,
Tzong-Wann Kao$^4$,
Yao-Hsin Chou$^1$}\vspace*{0.015truein}

\centerline{\footnotesize\it $^1$Department of Computer Science and
Information   Engineering}  
\baselineskip=10pt
\centerline{\footnotesize\it  National Taiwan University of Science
  and Technology, Taipei 10607, Taiwan}
\centerline{\footnotesize\it $^2$Department of Electronic Engineering, National United University}
\baselineskip=10pt
\centerline{\footnotesize\it Miao-Li 36003, Taiwan}
\centerline{\footnotesize\it $^3$Department of Physics, Center for Quantum Science and Engineering, and Center for
  Theoretical Sciences}
\baselineskip=10pt
\centerline{\footnotesize\it National Taiwan University, Taipei 10617, Taiwan}
\centerline{\footnotesize\it $^4$Department of Electronic Engineering, Technology and
  Science Institute of Northern Taiwan}
\baselineskip=10pt
\centerline{\footnotesize\it Taipei 11202, Taiwan}
\vspace*{0.225truein}
\publisher{December 20, 2007}{January 31, 2009}

\vspace*{0.21truein}

\abstracts{
In this paper, 
we first point out that some recently proposed quantum direct
communication (QDC) protocols with authentication are vulnerable
under some specific attacks, and the secrete message will leak out to 
the authenticator who is introduced to authenticate users
participating in the communication.
We then propose a new protocol that is capable
of achieving secure QDC with authentication as long as
the authenticator would do the authentication job faithfully.
Our quantum protocol introduces a mutual authentication procedure, uses 
the quantum Bell states, and applies unitary transformations in the authentication process.
Then it exploits and utilizes the entanglement swapping and local unitary operations 
in the communication processes. 
Thus, after the authentication process, 
the client users are left alone to communicate with each other, 
and the authenticator has no access to the secrete message. 
In addition, our protocol 
does not require a direct quantum link between any two users, 
who want to communicate with each other.
This may also be an appealing advantage in the
implementation of a practical quantum communication network.
}{}{}

\vspace*{10pt}

\vspace*{3pt}
\communicate{H-K Lo \& R Laflamme}

\vspace*{1pt}\textlineskip    
\section{Introduction}
\noindent
Quantum key distribution (QKD) is an approach using quantum mechanics
principles for the distribution of a secret key with unconditional
security \cite{m98,lc99,PhysRevLett.85.441}. 
Recently, there have been theoretical progresses and 
experimental demonstrations for the QKD protocols
\cite{PhysRevA.68.022317,PhysRevLett.91.087901,boileau:220501,ma:032330,zhao:070502,peng:010505}. 
Different from QKD, 
a quantum direct communication (QDC) protocol is to
transmit directly a secret message without generating in advance a secret
encryption key between the parties who want to communicate with each other.  
After the first proposal by Beige \textit{et al.} \cite{bekw02},   
many QDC protocols have been proposed \cite{PhysRevLett.89.187902,PhysRevA.68.042317,deng:052319,Man05,lucamarini:140501,wang:044305}.
But most QDC protocols are susceptible to the 
\textit{man-in-the-middle} (MITM) attack in which 
the eavesdropping attacker makes extra connections
with the victim users, and relays messages between them 
while making them believe
that they are talking directly to each other over a private connection.
In fact, the entire message communication is under control by the attacker. 
In order to prevent the MITM attack, several quantum authentication schemes 
have been put forward \cite{PhysRevA.64.062309,PhysRevA.60.149,PhysRevA.61.022303,PhysRevA.62.022305,PhysRevA.54.2651}.
Recently, Lee \textit{et al.} \cite{lee:042305} 
proposed two protocols which
combined QDC with user authentication. 
User authentication is to assure the communicating party is 
the one that he/she claims to be and 
the message is only communicated between the authentic users. 
This mechanism plays an important role in secure message communication 
against the MITM attacks.
However, Zhang \textit{et al.} \cite{zhang:026301} pointed out
that in the two protocols of Lee \textit{et al.}, the authenticator Trent who
is introduced to authenticate the users participating in the communication
should be prevented from knowing the secret message. 
They also showed that these two protocols are
vulnerable to some specific attacks by Trent.
To prevent the attacks, they revised
the original version of the protocols by using the Pauli $Z$ operation
$\sigma_z$ instead of the original bit-flip operation $X$ \cite{zhang:026301}.

In this paper, we first point out that 
the improved version of the protocols proposed by 
Zhang \textit{et al.} still cannot
prevent the authenticator Trent from knowing the secret
message if Trent would prepare different initial states.
To prevent both the authenticator Trent and an eavesdropper Eve
from knowing the secrete message,
we propose a new quantum protocol that is capable of achieving secure QDC 
as long as the authenticator Trent would
do the authentication job faithfully.
In our protocol, we introduce a mutual authentication procedure,
use the quantum Bell states instead of the GHZ states in 
\cite{lee:042305,zhang:026301},  
and apply the unitary transformations in the authentication process. 
Then we exploit and utilize the quantum entanglement
swapping and local unitary operations in the communication process.
In addition, our protocol which uses the beautiful feature of
quantum entanglement swapping does not require  
a direct quantum link between
any two clients/users who want to communicate with each other. 
This may also be an appealing advantage in the
implementation of a practical quantum communication network.

Similar to most of the proposed QDC protocols in the literature \cite{PhysRevLett.89.187902,PhysRevA.68.042317,deng:052319,Man05,lucamarini:140501,wang:044305,lee:042305,zhang:026301},
we present the proof-of-principle illustration of our 
secure QDC protocol against the attacks by
eavesdroppers, impostor users and authenticator. 
In a realistic implementation of a QDC protocol, there are many
other practical issues that need to be considered. For example, (i) the noise
(depolarization and dephasing) in the quantum communication channels,  
(ii) the imperfection of the Bell-state (EPR) source and distribution,
(iii) the errors that may occur during the quantum information storage, quantum
gate operations and quantum measurements, 
and (iv) the photon loss
inevitable in propagating light over distance through optical systems (if a
photonic implementation of the QDC protocol is adopted) are important 
problems that need to be dealt with.   
Issues similar to the first three mentioned above for our QDC protocol
have been discussed by 
Lo and Chau \cite{lc99} in the context of the security of QKD
over arbitrarily long distances.
They have shown that by combining the ideas of quantum
repeaters and fault-tolerant quantum computation, the security of QKD
in the presence of source, device, and channel noises as well as
operation and measurement errors could be made unconditionally
secure. 
As to make quantum state robust against photon loss, recently 
Wasilewski and Banaszek \cite{PhysRevA.75.042316} have proposed 
a three-photon quantum error correction code to protect an encoded 
qubit against a single-photon loss. 
They have also discussed the preparation of the
code as well as quantum state and process tomography in the code space 
using linear optics with single-photon sources and conditional detection.
We may apply the results of 
these studies \cite{lc99,PhysRevA.75.042316}(and references therein) to 
argue that our QDC protocol could also be made secure in a realistic setting
under similar conditions. 
However, an in-depth investigation to demonstrate that 
each of the practical issues mentioned above for our QDC protocol 
could be really resolved may be still required, 
but that is beyond the scope of this paper.
Nevertheless, we show here that our QDC protocol is secure against the
attacks by eavesdroppers, impostor users and authenticator.

\section{Attacks by the authenticator using different initial states}
\label{Sec:TrentAttacks}
\noindent
In order to introduce our mutual authentication process later and to
discuss our proposed attack by the authenticator Trent on the improved 
version proposed by Zhang \textit{et al.} \cite{zhang:026301},
we first summarize the authentication part in the protocol by 
Lee \textit{et al.} \cite{lee:042305} as follows. 

(1) The client users register their secret identities and one-way hash
functions with the authenticator Trent and then go apart. The user's
authentication key shared with Trent can be calculated as 
$h_{\rm user}(ID_{\rm user},C_{\rm user})$, where $ID_{\rm user}$ is
the user's secret identity sequence and
$C_{\rm user}$ is the counter of calls on the user's hash function,
$h_{\rm user}$. \cite{lee:042305}
 
(2) When Alice asks Trent that she would like to communicate with Bob, 
Trent generates N tripartite GHZ states \ket{\Psi} with  
$\ket{\psi_i}=\frac{1}{\sqrt{2}}(\ket{000}+\ket{111})_{ATB}$ and
$i=1,2,\cdots ,N$. The subscripts of A, T, and B correspond to Alice, Trent
and Bob, respectively.
Trent then makes unitary operations $I(H)$ on \ket{\Psi} according to the 
authentication key bit values $0(1)$ of Alice and Bob, respectively.

(3) Trent distributes the particles of $A$ sequence to Alice and the
    particles of $B$ sequence to Bob.

(4) Alice and Bob make reverse unitary operations on the received
    particles with their own authentication keys, respectively.

(5) After making local measurements in the $\sigma_z$ basis 
on a subset, Alice and Bob can compare the results through a classical public channel.

If the error rate is higher than expected (i.e., existence of an eavesdropper
in the communication), then Alice and Bob
terminate the protocol. Otherwise, they can confirm that their counter
parts are legitimate and the channel is secure.
Alice and Bob can then execute the message transmission 
procedures with Trent.
However, as it is pointed out by Zhang \textit{et al.}
\cite{zhang:026301}, the protocols by Lee \textit{et al.}
\cite{lee:042305} are vulnerable to the insider Trent with the \textit{intercept-measure-resend} attack, 
and therefore they proposed two improved schemes with different unitary
operations $H(HZ)$ instead of $I(H)$ on Lee \textit{et al.}'s protocols. 

We now show below the improved version of the two protocols proposed by 
Zhang \textit{et al.} \cite{zhang:026301} still cannot prevent the
attack from the authenticator Trent if he 
prepared different initial states from the states that he is supposed
to prepare.   
For example, if Trent wants to know Alice's secret message, 
he could prepare the initial state \ket{\Psi} as
$\frac{1}{\sqrt{2}}(\ket{+++}+\ket{---})_{ATB}$ instead of
$\frac{1}{\sqrt{2}}(\ket{000}+\ket{111})_{ATB}$, where \ket{\pm}
denotes $\frac{1}{\sqrt{2}}(\ket{0}\pm\ket{-})$ as usual and the
subscripts of A, T, and B indicate Alice, Trent and Bob's particles
(qubits), respectively. Then, after 
Alice's encoding operation $H_A$ or $H_{A}Z_{A}$ as in \cite{zhang:026301},
the state will 
become either $\frac{1}{\sqrt{2}}(\ket{0++}+\ket{1--})_{ATB}$ or
$\frac{1}{\sqrt{2}}(\ket{1++}+\ket{0--})_{ATB}$. If the
authentication is verified, Alice can send her qubits either to Bob
(protocol 1) or to Trent (protocol 2) \cite{zhang:026301}.
Then if Trent, just like the attacks proposed in \cite{zhang:026301}, 
intercepts ( protocol 1) or receives (protocol 2) 
Alice's qubit and measures it in the $\{0,1\}$ basis, he can then unambiguously
figure out the encoding operation and thus the bit value of Alice
after he has also measured his own qubit in the $\{+,-\}$
basis. In other words, if the measurement outcome is $(0,+)_{AT}$ or
$(1,-)_{AT}$, then Trent can conclude that Alice has performed a $H_A$
operation corresponding to the bit value of $0$. Otherwise, if the measurement
outcome is $(1,+)_{AT}$ or $(0,-)_{AT}$, then Alice has performed $H_A Z_A$
operation corresponding to the bit value of $1$. 
In this way, Trent can obtain Alice's whole
bit string including both the random bit string and the secret message. 
As Alice will publish the information regarding 
which qubits are used as the check qubits in public, Trent can then 
remove the random check bits.
As a consequence, Trent will have complete
knowledge of Alice's secret message.  After Trent's attack, he resends
Alice's qubit to Bob (protocol 1).
The different initial states might cause the error rate
higher than expected. Alice and Bob will, in this case, conclude that
there is an eavesdropper in the communication.  But they still think
the secret message has not leaked out. 
On the contrary, the secret message, in fact, has already leaked out to Trent.
So Trent can use the \textit{prepare-intercept-measure-resend} attack on the
improved scheme of protocol 1 or the \textit{prepare-measure} attack on the 
improved scheme of protocol 2 proposed by Zhang \textit{et al.} 
\cite{zhang:026301} to completely know the secret message. 

\section{Quantum secure direct communication with mutual
  authentication}
\label{Sec:QSDC}
\noindent
In this section, we present a new QDC
protocol that is able to prevent
all the specific attacks mentioned.
It may also seems that all the attacks mentioned above could be avoided if the
authenticator Trent is reliable. In fact, to the best of our knowledge,
nothing in the existing proposed protocols prevents an imposter to step in and pretend to
be the real authenticator Trent between a genuine user and unlawful
receiver. Our proposed protocol can, however, 
prevent such attack through mutual authentication.

In the protocols by Lee \textit{et al.} \cite{lee:042305} and by Zhang
\textit{et al.} \cite{zhang:026301} as well 
as in our protocol described later, Trent, as an authenticator,
is considered to be more powerful than the rest of other parties or users.
For example, all the user's secret identities are known to
the authenticator Trent, and all the quantum resources are issued by him.
It is thus important in the communication protocol that we
should first at least make sure whether ``Trent'' is the genuine 
authenticator or not. If not, some illegitimate party might pretend to
be Trent, then allow Alice to pass the authentication process, and finally
obtain the secret message. 
For instance, the imposter Trent might simply ask a fake Bob to
receive the secret message 
from Alice and get the secret message from the fake Bob later. 
We thus ask, in our protocol, the users/clients also to authenticate
Trent to prevent an imposter to step in and act as the authenticator.
Of course, if the real authenticator Trent would like to eavesdrop or steal 
the secret message, his role will then become similar to the imposter
Trent and he may also ask a fake Bob to receive
the secret message from Alice. 
This problem of a fake Bob 
could, for example, possibly be found (although not always perfectly
and immediately)  
by allowing the users/clients to access the classical public
channel at any time. If someone pretends to be Bob to communicate with Alice,
the real Bob may discover this event during the attack.

Nevertheless, if the real Trent will do his authentication job faithfully,
then it is desired that a scheme 
to prevent an imposter Trent from being able to manipulate the
authentication and communication processes, 
and to steal the secret message later is available.
We show below that our protocol with mutual authentication can
accomplish that goal and thus achieve secure QDC. 
The classical part of the generation of the authentication keys of the
users in our protocol is similar to that in Lee \textit{et al.}
\cite{lee:042305} 
when they registered to the authenticator Trent.  
The secret identity sequence, $ID$, and one-way hash function, $h$, 
of each user are known to Trent. 
For simplicity, we denote the authentication keys of Alice and Bob as 
$AK_{A}=h_{A}(ID_{A}, C_{A})$, $AK_{B}=h_{B}(ID_{B},C_{B})$,
respectively. $C_i$ is the counter of calls on the user's hash function, where $i=A,B$. 
If the length of $AK_i$, denoted as $l_i$, is not large enough
to cover the necessary operations, 
new authentication keys can be created by increasing the counter $C_i$ as
described in Ref.~\cite{lee:042305}. 
In order to secure the authentication process
to prevent an imposter authenticator to step in, 
we ask the user/client to authenticate Trent too.
It is thus a mutual authentication process.
In order for the users to be able to authenticate Trent,  
as well as to prevent Trent's different initial state attack, 
extra quantum resources, which include the introduction, manipulations and measurements of extra ancilla qubits, are issued by the users in our authentication protocol. 
Again, the reason why to authenticate each other in a mutual manner is 
to prevent the presence of an imposter authenticator and unlawful
users to steal
the secrete message.

One of the reasons why we use the Bell state instead of the GHZ state 
in our protocol is to improve the authentication process 
and to prevent the authenticator 
Trent from learning too much knowledge about the
secret communication.
We believe Trent's
responsibility is only to authenticate the users/clients who want to
communicate with other users/clients. As we will show later, after Trent
finishes his authentication job, he will be prevented from knowing
the secret message if the Bell state pairs and entanglement swapping
are employed. 
Another reason
why using the Bell state is that the two-particle Bell state can be used
in a peer to peer environment, while the three-particle GHZ state used in 
Refs.~\cite{lee:042305,zhang:026301} has
to involve with the third party's cooperation. That means if Trent wants to
authenticate Alice in the protocols of 
Refs.~\cite{lee:042305,zhang:026301}, he needs Bob's honest assistance if
the GHZ state is used. This will not be good 
as Alice's authentication has to depend on whether Bob is
honest or not.  In our authentication procedure the GHZ state is also
used, but this three-particle GHZ state is however a joint state
between Trent's qubit, Alice's qubit and an ancilla qubit. 
This ancilla qubit is introduced and controlled, for
example, by Trent when he wants to authenticate Alice.     

In addition, the process of the user
authentication should be closely connected to the message
communication process in order to protect against the attack of
\textit{modification}, \textit{delay}, \textit{replay}, and \textit{recording}  \cite{Schneier96} 
which could occur between these two processes.
We use the quantum entanglement swapping scheme
\cite{PhysRevLett.71.4287} with the Bell states to 
bridge these two processes. That means
the secret message will be transmitted only after the successful
authentication. 

The details of our protocol will be presented below. In Sec.~\ref{Sec:ES}, 
the important concept of entanglement swapping is discussed. The mutual
authentication process which uses the Bell-state entanglement
swapping and local operations is described in
Sec.~\ref{Sec:authentication}.  
The communication process which also employs the
entanglement swapping and local operations are described in
Sec.~\ref{Sec:communication}.

\subsection{The scheme of entanglement swapping} \label{Sec:ES}
\begin{figure}
\includegraphics[width=7cm]{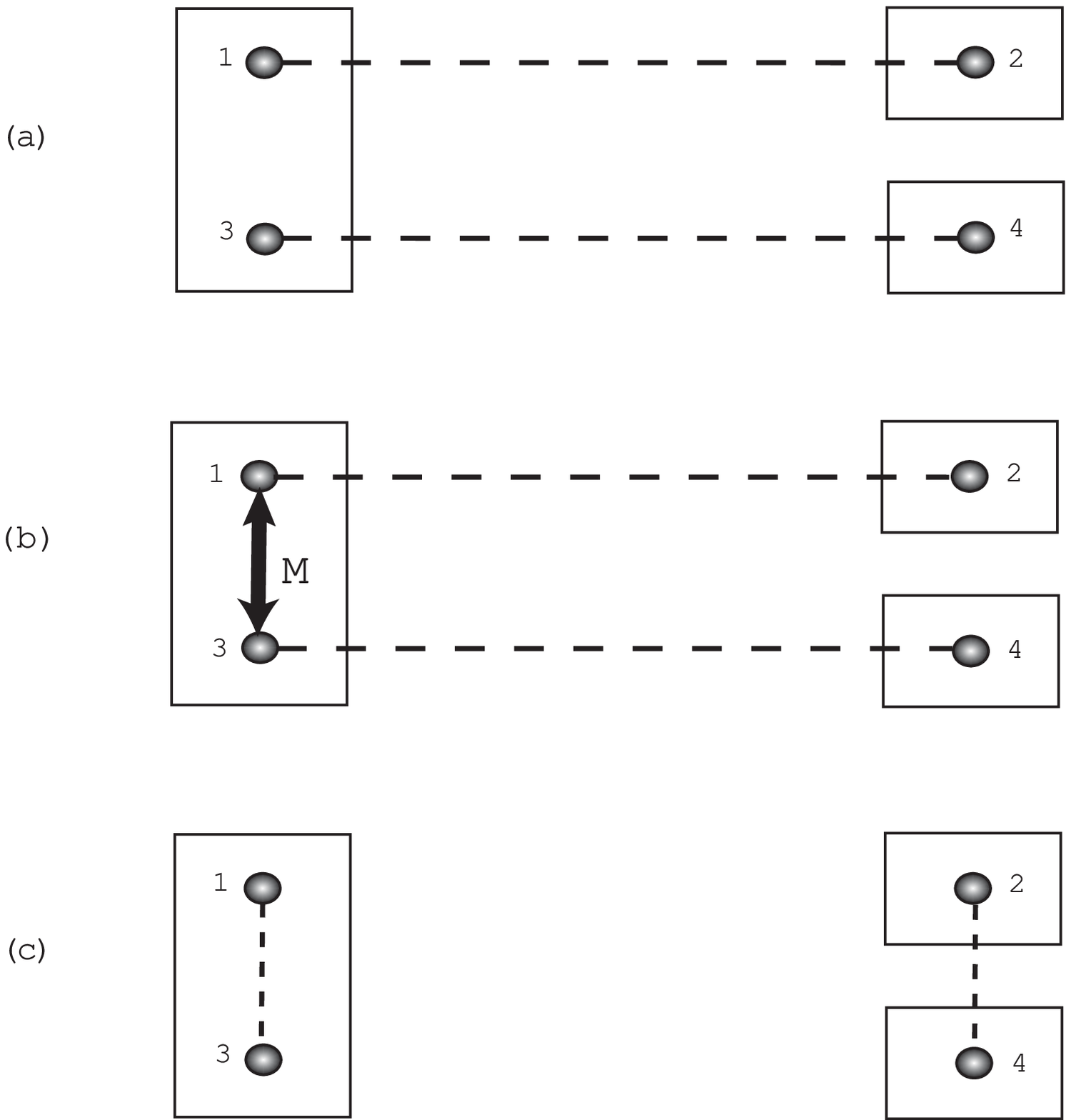}
\vspace{0.5cm}
\fcaption{\label{fig:fig1} Graphical illustration of the entanglement
  swapping scheme. (a) Two EPR pairs are shown in this figure, the dash line
  represents the entanglement shared between the two particles. (b)
  The action of Bell basis measurement is represented with M. (c) The
  measurement causes the entanglement swapping.} 
\end{figure}

\noindent
Entanglement swapping \cite{PhysRevLett.71.4287} is a method that
enables one to swap the entanglement of two entangled pairs of quantum
particles into that of two new pairs by local operations 
(see Fig.~\ref{fig:fig1}). The newly 
entangled particles may be spatially separately, without interaction
and without pre-shared entanglement between them.
To illustrate that, we first introduce four Bell states (or EPR pairs)
as follows: 
\begin{eqnarray}
\label{Bell_basis}
\ket{\phi^\pm}=\frac{1}{\sqrt{2}}(\ket{00}\pm\ket{11}),\nonumber\\
\ket{\psi^\pm}=\frac{1}{\sqrt{2}}(\ket{01}\pm\ket{10}).
\end{eqnarray}
The computational basis states, from Eq. (\ref{Bell_basis}), can be
expressed in terms of the four Bell basis states as:
\begin{eqnarray}
\label{Computational_basis}
\ket{00}=\frac{1}{\sqrt{2}}(\ket{\phi^+}+\ket{\phi^-}),\nonumber\\
\ket{11}=\frac{1}{\sqrt{2}}(\ket{\phi^+}-\ket{\phi^-}),\nonumber\\
\ket{01}=\frac{1}{\sqrt{2}}(\ket{\psi^+}+\ket{\psi^-}),\nonumber\\
\ket{10}=\frac{1}{\sqrt{2}}(\ket{\psi^+}-\ket{\psi^-}).
\end{eqnarray}
Table~\ref{tab:table1} illustrates the transformation between the Bell
state basis and the computational state basis.
\begin{table}
\tcaption{\label{tab:table1}Transformation table between the Bell state
  basis and computational state basis. This table not only illustrates
  \ket{\phi^+}=$\frac{1}{\sqrt{2}}$(\ket{00}+\ket{11}) or
  \ket{\psi^-}=$\frac{1}{\sqrt{2}}$(\ket{01}-\ket{10})
  columnwisely, but also illustrates
  \ket{00}=$\frac{1}{\sqrt{2}}$(\ket{\phi^+}+\ket{\phi^-}) or
  \ket{10}=$\frac{1}{\sqrt{2}}$(\ket{\psi^+}-\ket{\psi^-})
  rowwisely.} 
\centerline{\footnotesize\smalllineskip
\begin{tabular}{|c||c|c|c|c|} \hline
Basis  & \ket{\phi^+} & \ket{\phi^-} & \ket{\psi^+} & \ket{\psi^-}\\ \hline \hline
\ket{00}& $\frac{1}{\sqrt{2}}$ & $\frac{1}{\sqrt{2}}$ & 0 & 0\\ \hline
\ket{01}& 0 & 0 & $\frac{1}{\sqrt{2}}$ & $\frac{1}{\sqrt{2}}$\\ \hline
\ket{10}& 0 & 0 & $\frac{1}{\sqrt{2}}$ & $-\frac{1}{\sqrt{2}}$\\ \hline
\ket{11}& $\frac{1}{\sqrt{2}}$ & $-\frac{1}{\sqrt{2}}$ & 0 & 0\\ \hline
\end{tabular}}
\end{table}

Initially, if two parties each owns two particles, one in each of
the shared two entangled pairs 1-2 and 3-4 in \ket{\phi^+} states as
shown in Fig.~\ref{fig:fig1}(a),  
then the quantum state of the two Bell pairs can be rewritten as \cite{PhysRevA.57.822},
\begin{eqnarray}
\label{EPR_tensor}
&&\ket{\phi^+}_{12}\otimes\ket{\phi^+}_{34} \nonumber\\
&=&\frac{1}{2}(\ket{\phi^+}\ket{\phi^+}+\ket{\phi^-}\ket{\phi^-}+\ket{\psi^+}\ket{\psi^+}+\ket{\psi^-}\ket{\psi^-})_{1324}.
\end{eqnarray}

If the party who initially owned the particles 1 and 3 
makes a measurement in the Bell basis locally as illustrated in
Fig.~\ref{fig:fig1}(b), 
then the system  
will swap the entanglement pairs from 1-2, 3-4 into
1-3, 2-4 illustrated in Fig.~\ref{fig:fig1}(c). 
Depending on the measurement result, the resultant state is in
one of the four Bell pairs in Eq.~(\ref{EPR_tensor}) with equal
probability of $\frac{1}{4}$. The particles 2 and 4 would thus \cite{Man05,Man04} 
be entangled. Similar results can be obtained if the initially shared
entangled states are the Bell states other than \ket{\phi_+}.   
The most
significant feature of entanglement swapping is to enable one to
entangle two quantum systems (2 and 4) that do not have direct interaction
between them by a Bell basis measurement (on 1 and 3). 

\subsection{mutual authentication process}\label{Sec:authentication}
\begin{figure}
\includegraphics[width=7cm]{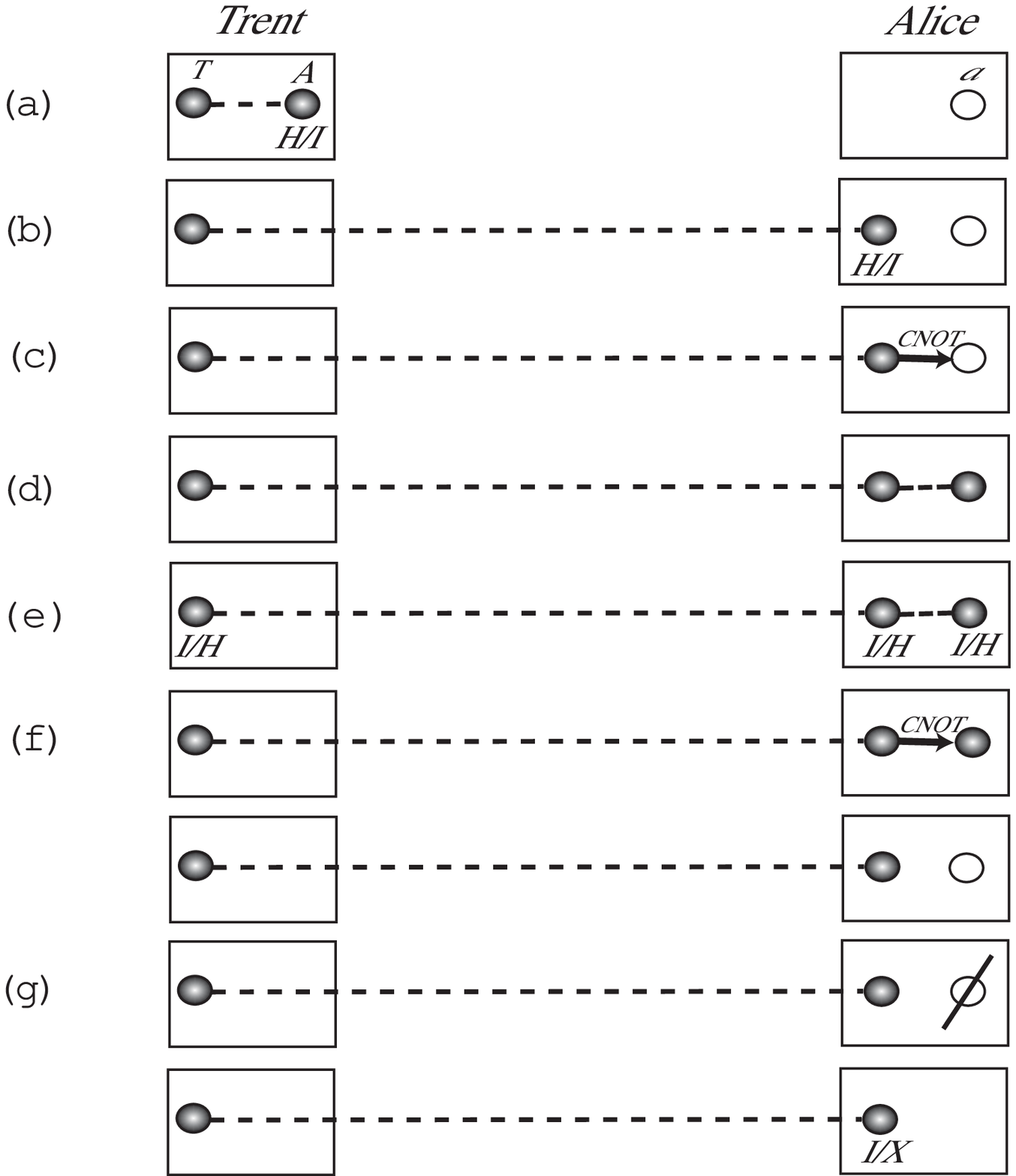}
\vspace{0.5cm}
\fcaption{\label{fig:fig2} Pictorial illustration of the authentication scheme. (a) Trent prepares an ordered $N+2v$ two-particle Bell states $\ket{\phi^+}=\frac{1}{\sqrt{2}}(\ket{00}+\ket{11})_{TA}$, and encodes each of the particles of the $A$ sequence by operation $H$ ($I$) according to the bit value of $0$ ($1$) of $AK_A$, here only one pair in verifying set $V$ and Alice's ancilla particle are shown. (b) Trent sends the particle $A$ to Alice, Alice then decodes the particle by operation $H$ ($I$) also according to the bit value of $0$ ($1$) of $AK_A$. (c) Alice uses the decoded particle to make the \textup{CNOT} operation on the ancilla. (d) The ancilla particle will be entangled with the original two particles. (e) Alice and Trent perform the operation $I$ ($H$) on their own particles according to the bit value of  $0$ ($1$) of $AK_A$ separately. (f) The Steps of recovering is also depended on the bit value of $AK_A$. If the bit value of $AK_A$ is $0$, Alice makes the operation \textup{CNOT} on the ancilla $a$ again. The ancilla will leave the entanglement after the operation \textup{CNOT}. (g) If the bit value of $AK_A$ is $1$, Alice measures the ancilla $a$. Alice then makes the operation $I$ ($X$) on the particle $A$ according to the measurement result of ancilla being $0$ ($1$).
}
\end{figure}

\noindent
After sharing the authentication keys with Trent, respectively, 
as described in Sec.~\ref{Sec:QSDC}, 
the clients/users, say Alice and Bob, might leave Trent and go apart. 
Suppose neither Alice nor Bob can see each other in a network environment.
When Alice wants to communicate with Bob, 
she and Bob must go through the authentication process with Trent first.
We now introduce the scheme of quantum \textit{Controlled-NOT} operations 
as well as local operations into our mutual authentication process. 
This authentication process
is illustrated schematically in Fig.~\ref{fig:fig2} 
and is described in detail as follows. 

(1) Once Trent receives Alice's request, he prepares an ordered $N+2v$
    two-particle Bell states, each of which is in, for example, the state 
    $\ket{\phi^+}=\frac{1}{\sqrt{2}}(\ket{00}+\ket{11})_{TA}$, where $N$
    is the number of secret message bits that is intended to be transmitted 
    in this round and $v$ is a
    sufficient large number for checking the noise of the quantum
    channel. In general, $v \geq l_A$, where $l_A$ is the length of
		Alice's authentication key, $AK_A$. 
		Trent keeps one of the particles (qubits) in each of the ordered $N+2v$
		two-particle Bell pairs  
		to form an ordered particle (qubit)
		sequence, called the $T$ sequence. 
		He then encodes each particle of the other correspondingly ordered sequence, called the $A$ sequence, 
		by an operation $H$ or $I$ according to the bit value of $AK_A$ being $1$ or $0$, respectively. 
		That is , if the key bit value is $0$, a $H$ operation is performed on the particle; otherwise, nothing is done.
		Since the number $N+2v$ is normally larger than the length  $l_A$ of $AK_A$, the key bits of $AK_A$ will be reused from 			the beginning for encoding.
		The process will be repeated until all $N+2v$ particles in the $A$ sequence are encoded with corresponding $H/I$ 							operations. Trent then sends the encoded particles of the ordered $A$ sequence to Alice. 

(2) After receiving the $A$ sequence, Alice decodes each of the particles 
    by an operations $H$ or $I$ 
    also according to the bit value $0$ or $1$ of $AK_A$ as Trent did. 
    The state of each of the ordered $N+2v$
    two-particle Bell pairs will return to its initial state, 
    $\frac{1}{\sqrt{2}}(\ket{00}+\ket{11})_{TA}$. 
    She then chooses randomly a
    sufficient large subset of size $v$ from the $A$ sequence, and pairs each
    of the chosen particle with its corresponding counterpart in the $T$ 
    sequence into a verifying set $V$. 
    One pair of particles in set $V$ is shown in
    Fig.~\ref{fig:fig2} (a) and (b). She also prepares $v$ ancilla
    particles, called the $\it{a}$ sequence, each of which is in the state
    \ket{0}. She uses the particles which she owns in set $V$ to be
    the control qubits and makes the quantum \textit{Controlled-NOT}
    operations (\textup{CNOT}) on the target qubits of the ancilla $a$ 
    in sequence. Then, as
    shown in Fig.~\ref{fig:fig2} (c) and (d), each of the ancilla
    particles (qubits) will be entangled with the original two particles 
    (qubits) in
    the Bell state. These three particles will then be in the state
    $\frac{1}{\sqrt{2}}(\ket{000}+\ket{111})_{TAa}$, where the
    subscripts of $T$, $A$ and $a$ denote the Trent, Alice and ancilla
    particles, respectively. Note that this state is just the quantum
    GHZ state used in the protocols of Lee \textit{et al.} \cite{lee:042305} 
    and by Zhang \textit{et al.} \cite{zhang:026301}.  The difference
    is that here the third particle, the ancilla qubit, is introduced and
    controlled by Alice when she is about to authenticate Trent. 

(3) Alice goes to identify Trent in case that some illegitimate party might 
    pretend to be the authenticator Trent or Trent might prepare different initial states to steal message. 
    She makes operation $I$ ($H$) on each of her two own particles 
    according to the bit value $0$ ($1$) of $AK_A$. 
    For example, if the \textit{i}th bit value of $AK_A$ is $0$, Alice performs
    the identity operation $I$ on each of her two particles in the \textit{i}th
    position in set $V$. Then, the three particles would remain in
    $\frac{1}{\sqrt{2}}(\ket{000}+\ket{111})_{TAa}$. On the contrary,
    if the \textit{i}th bit of $AK_A$ is $1$, she performs the
    operation $H$ on each of her two particles 
    and the three particles would become
    $\frac{1}{2\sqrt{2}}(\ket{0(0+1)(0+1)}+\ket{1(0-1)(0-1)})_{TAa}$. 
   
(4) Alice tells Trent the positions of the set $V$ particles in the original
    $N+2v$ sequence and tells him that she has finished the 
    transformation operations over a classical public channel. 
		Trent then also makes operation $I$ ($H$) on his
    own particle according to the bit value of his own key $AK_A$ in
    set $V$ in sequence. For example, if the \textit{i}th bit of
    $AK_A$ is $0$, the three particles would still stay in
    $\frac{1}{\sqrt{2}}(\ket{000}+\ket{111})_{TAa}$. Otherwise, if the
    \textit{i}th bit of $AK_A$ is $1$, as
    illustrated in Fig.~\ref{fig:fig2}(e), the three particles would
    become 
\begin{eqnarray}
&&(H\otimes H\otimes H)_{TAa}\frac{1}{\sqrt{2}}(\ket{000}+\ket{111})_{TAa}\nonumber\\
&&\Rightarrow\frac{1}{2}(\ket{(00+11)0}+\ket{(10+01)1})_{TAa}.
\end{eqnarray}

(5) After Trent tells Alice that he has finished his operations 
    over a public channel, Alice
    will perform different operations for each pair in set $V$
    according to the \textit{i}th bit value of $AK_A$. If the
    \textit{i}th bit value of $AK_A$ is $0$, she will make the
    operation \textup{CNOT} on the \textit{i}th ancilla $a$ again with
    particle $A$ as the control qubit. Then the ancilla qubit will lost
    the entanglement with the pair of qubits $T$ and $A$ as shown in
    Fig.~\ref{fig:fig2}(f). On the contrary, if the \textit{i}th bit
    value of $AK_A$ is $1$, Alice will measure the \textit{i}th
    ancilla particle $a$ in the computational \{$0$,$1$\} basis, i.e.,
    the $\sigma_Z$ basis. If there is no eavesdropping or
    interference, Alice will obtain either $0$ or $1$ with equal probability
    $1/2$. If she obtains $1$, she makes the operation $X$ on the particle
    $A$. Otherwise, nothing is done. These actions are shown in
    Fig.~\ref{fig:fig2}(g). After this, 
    the pair of the two particles $T$
    and $A$ in set $V$ should return back into the original Bell state
    after consuming the ancilla, provided that there is no
    eavesdropper, Eve, present. 

(6) Now, Alice is going to authenticate Trent as well as to check the
    presence of Eve. After Alice measures her own particles (qubits) in
    set $V$ one by one in the $\sigma_Z$ basis, she informs Trent
    that her measurements are finished 
    (but does not reveal her measurement results) in public.
    Trent then also measures his own particles in the $\sigma_Z$
    basis and tells Alice the results in public. At last, Alice compares her
    results with those of Trent's to authenticate Trent. If they
    have a sufficient large number of results that are the same, Alice \cite{Man05,Man04} accepts
    that Trent is the real Trent (the authenticator) and she proceeds
    the steps to 
    be authenticated by Trent. Otherwise, if the error rate is too
    high, she just stops the procedure. 

(7) Next, Trent will authenticate Alice. 
		This reverse authentication procedure 
		could be much simpler as compared 
		with the above steps. This is because Trent owns the particle he prepared in the very beginning and does not need to 
		check possible different initial state attack by Alice, so no extra ancilla qubits need to be introduced. 
		If Alice can decode the $A$ sequence in Step 2, the remaining $N+v$ pairs will all return to the initial state, 			 				$\frac{1}{\sqrt{2}}(\ket{00}+\ket{11})_{TA}$. Trent then randomly selects $v$ particles in the remaining $T$ sequence and 		pairs with Alice's particles in $A$ sequence to form the reverse verifying set $V'$.
		He then measures his own particles (qubits) in set $V'$ one by one in the $\sigma_Z$ basis, and informs Alice
    that her measurements are finished in public.
    Alice then also measures his own particles in the $\sigma_Z$
    basis and tells Trent the results in public. Trent then compares his
    results with those of Alice's to authenticate Alice 
    and also check the existence of Eve. If their measurement results agree
    with a sufficiently high probability, Trent \cite{Man05,Man04} accepts
    that Alice is a legitimate user/client.  
    Otherwise, if the error rate is too
    high, he just stops the procedure. 
		This is exactly the reverse process in Step 6 with the interchange of the roles of Alice and Trent.

(8) After the mutual authentication is finished, there
    still are $N$ pairs of the Bell state
    $\frac{1}{\sqrt{2}}{(\ket{00}+\ket{11})_{TA}}$ between Trent and
    Alice's qubits. 
    Note that the local operations after the measurements of the ancilla qubits 
    in Steps 5 depend on the initially
    prepared Bell state. In the above example, the initial and recovered
    Bell state is 
    $\frac{1}{\sqrt{2}}{(\ket{00}+\ket{11})_{TA}}$. Other
    Bell states can also be used in our protocol with a
    slight modification of the local operations. Their relations
    are illustrated in Table~\ref{tab:table2}.
    Since both Alice and Trent choose the verifying sets at random,
    they could also check the security of the channel during the
    authentication steps. Not only the illegitimate party but also the
    existence of the eavesdropper Eve could be detected during the
    verification process. Furthermore, the message communication process
    will proceed only if the authentication process is
    successful. If the channel is too noisy with a high error rate, 
    they would stop the procedure and start over again. 

\begin{table}
\tcaption{\label{tab:table2}Relations between the initially prepared
    Bell states, the measurement result of ancilla, and the recovering
    operations on the home qubit when the bit value of $AK$ of the
    other party is $1$.}
\centerline{\footnotesize\smalllineskip 
\begin{tabular}{|c||c|c|} \hline
Initial prepared & \multicolumn{2}{c|}{Recovering operations on the self qubit}\\ \cline{2-3}
Bell States  & ancilla's outcome: $0$   & ancilla's outcome: $1$\\ \hline \hline
\ket{\phi^+}=$\frac{1}{\sqrt{2}}$(\ket{00}+\ket{11}) & $I$ & $X$ \\ \hline
\ket{\phi^-}=$\frac{1}{\sqrt{2}}$(\ket{00}-\ket{11}) & $X$ & $I$ \\ \hline
\ket{\psi^+}=$\frac{1}{\sqrt{2}}$(\ket{01}+\ket{10}) & $iY$ & $Z$ \\ \hline
\ket{\psi^-}=$\frac{1}{\sqrt{2}}$(\ket{01}-\ket{10}) & $Z$ & $iY$ \\ \hline
\end{tabular}}
\end{table}

(9) Finally, Trent notifies Bob that Alice wants to communicate with
    him. Likewise, Bob and Trent can authenticate each other. If
    nothing goes wrong, they will also keep $N$ pairs of the Bell state
    $\frac{1}{\sqrt{2}}{(\ket{00}+\ket{11})_{TB}}$.  

\subsection{Communication process}\label{Sec:communication}
\noindent
After finishing the authentication process, Trent's qubit is entangled
with Alice's qubit in the Bell state $\frac{1}{\sqrt{2}}(\ket{00}+\ket{11})$
and likewise also with Bob. We describe below a \textit{session-key} based
communication process \cite{Schneier96} which uses the skills of
entanglement swapping.    
Here, the \textit{session key} indicates that 
the sequence of the entangled states between Alice and Bob's qubits, 
generated as a result of the Trent's entanglement swapping measurements,
is used for only one particular communication session. The detailed
communication process is described as follows.

(1) Trent makes a Bell measurement on his own two qubits, one
    particle entangled with Alice's qubit and the other entangled
    with Bob's qubit, in each of the $N$ Bell pairs in
    sequence. Each time Trent will obtain a result with equal.
    probability $1/4$ out of four possible outcomes corresponding to
    the resultant four possible Bell states in which Alice's qubit 
    and Bob's qubit
    will be entangled. In other words, as a consequence of Trent's
    Bell measurement, the entanglement has been swapped into the joint state
    of Alice and Bob's qubits. However, Alice and Bob
    do not know exactly which entangled Bell state their qubits 
    really share so far. 
    Note that there is no qubit (particle) transmitted between Alice
    and Bob, so the eavesdropper
    Eve cannot obtain any quantum information during this process. 

(2) Trent announces the results of his Bell measurements in sequence
    over a classical public channel. 
    Another appealing feature as a result of the entanglement
    swapping is that Trent will leave the clients, Alice and Bob, alone to
    communicate with each other. In other words, after the 
    authenticator Trent finishes his job, he will not be involved in 
    the message communication process and thus he is 
    prevented from learning the secret message.

(3) After Trent's public announcement, Alice and Bob then have the knowledge
    about the identity of each of the shared Bell state between their
    qubits in the sequence. 
    They can use the sequence of the shared entangled Bell pairs
    to send the secret message 
    using the following two different schemes. 
    
    (i) They may use the scheme of \textit{dense coding}, first
    proposed by Bennett and Wiesner \cite{PhysRevLett.69.2881}, to
    transmit two classical bits of information using one entangled Bell
    pair. As Alice and Bob know which Bell state they
    share, say in $\ket{\phi_+}=\frac{1}{\sqrt{2}}(\ket{00}+\ket{11})$,
    Alice can 
    then determine and encode the two classical bits that she wants to
    send to Bob ($00$,$01$,$10$ and $11$) into the 
    unitary operation ($U$=$I$,$X$,$Z$ or $iY$) performed on her
    particle (qubit) of the Bell pair. Alice then sends her particle to Bob.
    After Bob receives Alice's particle, Bob can measure the two particles 
    in the Bell basis and 
    obtain Alice's encoded information. At the end of the direct 
    communication process, Alice can send $2N$ bits of classical
    information to Bob with these remaining $N$ entangled Bell pairs.
    However, since Alice's particles are transmitted to Bob, Eve might
    intercept them in the middle. To guarantee secure communication,
    some randomly chosen entangled Bell pairs have to be used 
    to check if Eve is eavesdropping \cite{PhysRevA.68.042317,wang:044305}.
    
    (ii) As mentioned, the above dense coding 
    scheme requires that there is a quantum channel between 
    Alice and Bob so that Alice can send her qubit to Bob. This,
    however, may not be practical in a realistic quantum communication
    network, as a direct quantum 
    link is required to be established between every two
    users/clients who want to communicate with each other. 
    To overcome this problem, we use a scheme \cite{Man05,Man04}, which
    also utilizes 
    the entanglement swapping together with local quantum operations, 
    to encode and transmit the message. 
    This scheme, proposed in the encoding-decoding step in 
    Refs.\cite{Man05,Man04}, does not need the qubits to be
    transmitted between Alice and Bob, and thus no
    quantum channel is required between them. 
    It, however, uses two
    entangled Bell pairs between Alice and Bob to transmitted two bits of
    information.  
    For example, in this scheme in Refs.\cite{Man05,Man04}, 
    Alice and Bob agree to apply one of the
    4 different unitary operations (say $U$=$I$,$Z$,$X$ or $iY$) on one
    particle of the two entangled Bell pairs to encode one of the 4
    different 2-bit 
    messages (say $00$,$01$,$10$ or $11$). 
    Suppose that the state of the two entangled pairs, 1-2 and 3-4, is
    the state of Eq.~(\ref{EPR_tensor}). 
    After Alice applies one of the local unitary operations, say $Z$, 
    on one of her two own particles, say particle 1 in
    Eq.~(\ref{EPR_tensor}), according to her 
    bit string values of $01$, the state becomes
\begin{eqnarray}
\label{Bell_communicate}
&&Z_1(\ket{\phi^+}_{12}\otimes\ket{\phi^+}_{34}) \nonumber\\
&=&\frac{1}{2}(\ket{\phi^-}\ket{\phi^+}+\ket{\phi^+}\ket{\phi^-}+\ket{\psi^-}\ket{\psi^+}+\ket{\psi^+}\ket{\psi^-})_{1324}.
\end{eqnarray}
    Alice can perform a Bell state basis measurement on her two
    particles, say particle 1 and 3, of the two entangled Bell pairs.  
    This then results in the entanglement swapping to the
    pairs of 1-3 and 2-4. 
    She then announces the measurement result, say in \ket{\psi^-} state, 
    to Bob through a classical public channel. Then after Alice's Bell
    measurement, Bob should obtain
    \ket{\psi^+} by his Bell basis measurement.
    Bob can read out Alice's bit string value $01$ after comparing the
    results of his own Bell basis measurement with Alice's.
    This scheme takes advantage of entanglement swapping. As a result, 
    it does not require a quantum channel between Alice and Bob and it 
    also avoids the possible eavesdropper 
    gaining any meaningful information of the secret message 
    during the communication process.
    Note that the resultant shared entangled Bell states between 
    Alice and Bob's qubits in our protocol are dependent on 
    the Trent's Bell measurement results. 
    Thus the entangled Bell states might not be all the same as 
    those used in our example or in Refs.\cite{Man05,Man04}.
    This, however, is not a problem as using two entangled pairs with
    different Bell states can also do the job if they know what the
    states they share 
    \cite{PhysRevA.57.822}.  
		Repeat the procedure in sequence, $N$ entangled Bell pairs can 
		transmit only $N$ bits of information.
		But this may not be a disadvantage since to 
		check and guarantee no eavesdropping in the dense coding scheme, 
		the consumption of entangled
    pairs may be large \cite{Man05} and might even be larger than $N/2$. 

\section{Discussions and security analysis}\label{Sec:discussion}
Before we conclude, a remark between our protocol and a QKD-like scheme 
as well as a security analysis of our protocol are in place.

\subsection{Comparison with a simple QKD scheme}\label{Sec:comparison} 
If the whole point of our protocol up to step (3) of Sec.~\ref{Sec:communication}, i.e., 
entanglement swapping after successful mutual authentication, 
is to certify only that Alice and Bob, the two users who want to communicate with 
each other, share perfect EPR pairs, then one may think that a simpler way, 
as is used in QKD, may be employed to do the same job. For example, 
if Alice would like to communicate with Bob in an EPR-pair-based QKD scheme for QDC, 
she may need (or ask someone else) to generate EPR pairs,
then transmits halves of the EPR pair qubits to Bob
and keeps the other halves to herself. 
Alice and Bob measure and then compare over a classical authenticated
channel the randomly selected $X$- or $Z$-basis measurement results on
some randomly chosen EPR pairs they share originally. If their
measurement results and their authentication keys (secrets) agree,
then they are certain that the remaining pairs they share are perfect
EPR pairs and that the other party is the real Alice or the real Bob. 
Otherwise, they may conclude that the quantum channel Alice used to
transmit qubits to Bob was too noisy, or that an eavesdropper Eve has
interfered in the qubit transmission process. 
Indeed, this QKD-like scheme combining with secure, authenticated
classical channels can certify that Alice and Bob share perfect EPR
pairs. 
So the absolute security of the classical authenticated channels must
be guaranteed for this QKD-like scheme for QDC to work. 
Suppose there is a shared secret key beforehand between the two users, Alice and Bob. They may
apply the classical Wegman-Carter scheme \cite{W-C:1981} for
authentication and for the comparison of the measurement results. For
example, they can use
the shared secret key to create Wegman-Carter tags and then compare
the hash values computed from the tags and the message that contains the
measurement results. The Wegman-Carter authentication
scheme \cite{W-C:1981} is unconditionally secure provided that 
the shared secret key bits used to create the tags are different each time.
But if Alice and Bob would like to authenticate each other
again for another communication, 
then the shared secret
key bits used to create the tags 
will be gradually used up in the Wegman-Carter scheme \cite{bb84}.
It was pointed out in Ref.~\cite{bb84} that the secret
key bits cannot be reused  
without compromising the provable security of the Wegman-Carter 
authentication scheme \cite{W-C:1981}.
So if no further process to replace or refresh the secret key bits,
then the provable security of the Wegman-Carter 
authentication scheme \cite{W-C:1981} may concede. 
One may use quantum channels to transmit new secret key bits
as is done in QKD. But if a secret
encryption key needs to be generated each time in advance 
between the parties who want to communicate and authenticate with each other, 
then this QKD-like scheme is similar to  
formal QKD rather than QDC that is intended here.  

In addition, in the QKD-like scheme each user needs to generate EPR pairs for every other users or a third party, say Trent, 
should be asked to prepare and distribute the EPR pairs for every users. 
But if Trent does not play also the role as an authenticator, 
then any two users have to authenticate and compare the measurement results directly 
between themselves through authenticated classical channels. 
As a result, each user needs to share a different secret key with 
every other user, 
and an authenticated classical channel is required 
between any two users who want to communicate with each other. 
Furthermore, if there is a new client, say Charlie, wants to join this communication network, 
his shared secret key needs to be generated and distributed securely between him and the rest of every client user.
These may not be
practical in the implementation of a realistic quantum communication
network as there may be many users in the network and 
they may be spatially far apart. 
These are the reasons why in the protocols of
Refs.~\cite{lee:042305,zhang:026301} as well as in our protocol, an
authenticator Trent is introduced in the QDC network. 
Thus one should consider
applying the QKD-like scheme to the similar protocols with an
authenticator Trent.  

Compared with the QKD-like scheme, 
there is, however,  no classical authenticated channel used
in our protocol. The classical channels used in our protocol are
public channels. They are not used to authenticate but are used to
broadcast (exchange) the classical information and measurement results
between the participants in public, as are used in
Refs.~\cite{lee:042305,zhang:026301}. The generation and registration 
of the classical
authentication keys of the users by the authenticator, as similar to that in
Refs.~\cite{lee:042305,zhang:026301}, does not mean the classical
authenticated channels are used. Since the users will go apart after
getting their authentication keys respectively and since no further
encryption scheme is used in the classical public channels in our
protocol, the users and authenticator cannot securely authenticate
each other through the classical public channels remotely. The
classical authentication keys of the participants are, however,
encoded with local quantum operations $H/I$ onto the EPR pairs of the
verifying sets as illustrated in Sec.~\ref{Sec:authentication}. 
Our authentication scheme is based on quantum entanglement, 
quantum operations and the randomness
of quantum measurement results.
So the presence of Eve will be discovered, and no useful information about the secret authentication key may be inferred in our protocol at least for the several possible Eve's attacks presented in Sec.~\ref{Sec:Security-analysis}.
Furthermore, our protocol can also avoid Trent's different 
initial states attack that the protocols in Refs.~\cite{lee:042305,zhang:026301} fail to prevent (see Sec.~\ref{Sec:TrentAttacks}), as extra quantum resources, which include the introduction, manipulations and measurements of extra ancilla qubits, are issued by the users in our authentication protocol when the users authenticate the authenticator Trent.
Next, we perform a security analysis of our protocol 
and show that this is the case.

\subsection{Security analysis}\label{Sec:Security-analysis}
Since after successful mutual authentication process, 
our protocol utilizes entanglement swapping and local quantum operations in the communication process.  As a result, 
it does not require a direct quantum link between any two users who want to communicate with each other and thus it 
also avoids the possible eavesdropper 
gaining any meaningful information of the secret message 
during the communication process.
We therefore focus the security analysis 
only on the authentication process. 
For simplicity, we assume Trent prepares the initial EPR pairs in the state
$\ket{\phi^+}_{TA}=\frac{1}{\sqrt{2}}(\ket{00}+\ket{11})_{TA}$ 
as before. 

First, if there is no eavesdropping and no other interference,
the resultant state, 
after the $H/I$, $H/I$, CNOT, $I/H$ and $I/X$ operations according to the corresponding bit value of the authentication key and the local measurement result of ancilla qubit,
should return back into its initial state $\ket{\phi^+}_{TA}$   
as illustrated in Sec.~\ref{Sec:authentication}.
Alice and Trent can then authenticate each other and detect the existence of Eve by comparing the 
$Z$-basis measurement results of their respective qubits in this EPR state.
If these measurement results agree, then they are sure that 
the opposite party really owns the pre-issued authentication key and holds halves of the EPR pairs.
An important observation of our authentication scheme 
is that no matter what the bit value 
of the authentication key is, the pair of the two particles $T$
and $A$ in the verifying sets $V$ return back into the 
original Bell state, $\frac{1}{\sqrt{2}}(\ket{00}+\ket{11})_{TA}$,
after consuming the ancilla as described 
in Step (5) of Sec.~\ref{Sec:authentication}. 
So the $Z$-basis measurement result of either $0$ or $1$ of particle $T$ 
obtained with equal probability and then announced in public 
by Trent reveals no useful information of the secret bit value of the 
authentication key, $AK_A$. 
Similarly, the state of each pair in the verifying set $V'$ is  also back to  
$\frac{1}{\sqrt{2}}(\ket{00}+\ket{11})_{TA}$ 
and thus no information on the secret bit values of Alice's key 
can be inferred from
the public announcement of Alice's qubit measurement results 
when Trent authenticates Alice.

Second, Eve may use {\it intercept attack}, that is, Eve intercepts the EPR particles sending to Alice, pretends herself to be the legitimate user Alice and tries to cheat Trent into an acceptance of her as Alice during the authentication process.
Eve who did not know Alice's key though
can, while authenticating Trent, just (pretend to be Alice to) accept Trent's measurement results announced in public to pass the process. 
However, in the reverse process in which Trent authenticates Eve 
(the fake Alice),  
Eve who does not know Alice's key cannot decode back the encoded qubit 
and thus cannot escape from the check by Trent as there will be a high error 
rate occurred when Trent compares Eve's qubit measurement results 
with his in the verifying set $V'$ of the checking step. 
In addition, we show below that Eve also cannot obtain any useful information about Alice's secrete key $AK_A$.
When the authentication process starts, Trent follows the protocol to perform an operation $H/I$ on each of the particles in the sequence $A$ according to the bit value of $0/1$ of $AK_A$. Suppose that Eve does nothing (she may do any operation but that will not affect the main conclusion of the following analysis) as she has no idea about the bit value of $AK_A$. The resultant state after Trent's next operation $I/H$ operation is $\frac{1}{\sqrt{2}}(\ket{0+}+\ket{1-})_{TE}$ if the bit value of $AK_A$ is $0$, and is $\frac{1}{\sqrt{2}}(\ket{+0}+\ket{-1})_{TE}=\frac{1}{\sqrt{2}}(\ket{0+}+\ket{1-})_{TE}$ if the bit value of $AK_A$ is $1$, where $\ket{\pm}=\frac{1}{\sqrt{2}}(\ket{0}\pm\ket{1})$ are the eigenstates of the $X$ operator with eigenvalues $\pm$. 
Note that Trent's later $I/H$ operation on his particle in the sequence $T$ according to the bit value $0/1$ of $AK_A$ is opposite to his first encoding operation. The encoding operation of $I/H$ on the particle in the sequence $A$ is, however, according to the bit value of $1/0$. These operations make 
Trent's resultant $Z$-basis qubit measurement results with equal probability of being either $0$ or $1$ independent of the bit value of $0/1$ of $AK_A$.   
So if Trent then follows the protocol to announce the $Z$-basis measurement result of his qubit (particle) in the verifying set $V$ one by one, then no matter what the bit value of $AK_A$ is, his measurement result will half-chance be $0$ and half-chance be $1$.   
Another case is the {\it intercept-and-CNOT attack}.
That is, if in the beginning, Eve also introduces an ancilla qubit $E_a$ being in $\ket{0}$ state and performs a CNOT operation on the intercepted qubit $E$ and the ancilla qubit, then the resultant state after Trent's $I/H$ operation are both in $\frac{1}{\sqrt{2}}[\ket{0}(\ket{00}+\ket{11})+\ket{1}(\ket{00-11})]_{TEE_a}$, no matter what the bit value of $AK_A$ is. Similar to the above scenario, regardless of Eve's subsequent operations, Trent will announce his $Z$-basis qubit measurement results with equal probability of being either $0$ or $1$, no matter what the bit value of $AK_A$ is. So no information of the secret key is revealed by Eve's intercept attack in both of the above cases.
In the reverse authentication process, since Eve does not know Alice's key, she cannot decode back the original EPR state. 
Suppose again Eve does nothing (she may do any operation but that will not affect the main conclusion of the following analysis). 
The resultant pair state in the verifying set $V'$ will be $\frac{1}{\sqrt{2}}(\ket{0+}+\ket{1-})_{TE}$ if the bit value of $AK_A$ is $0$, and is $\frac{1}{\sqrt{2}}(\ket{00}+\ket{11})_{TE}$ if the bit value of $AK_A$ is $1$.
So the fake Alice's (Eve's) qubit measurement result with equal probability of being either $0$ or $1$ cannot infer useful information about the real Alice's authentication key. It is obvious to see that Eve may do any operation instead of doing nothing on her qubits, but her measurement results will have no relation at all with the Alice's key. 
So Eve's {\it intercept attack} can catch nothing except being discovered.
  
Third, Eve could use {\it intercept-and-resend attack}, i.e.,
Eve first intercepts Trent's EPR particle $A$ sent toward Alice,
and then transmits the particle $E_A$ of the EPR pair that she prepared
to Alice instead.
Eve keeps particles $A$ and $E$ in her hands, which are entangled, respectively, with the Trent's and Alice's particles.
Eve may also try to first prepares an additional ancilla qubit in the  $\ket{0}$ state and entangles it with Trent's EPR state or with her prepared EPR pair by CNOT operation.
Without knowing Alice's authentication key, Eve's attack cannot pass Trent's authentication as stated above. In addition, Eve will again obtain no information of the secrete key bit when she try to authenticates Trent or authenticate Alice with the similar reasons stated also above. As a result, Eve's {\it intercept-and-resend attack} will also fail, and Eve will not get any useful information of the secret keys, either.

From the above analysis, Eve's several possible attacks will be discovered 
during our authentication process and 
furthermore, Eve cannot infer useful information about 
Alice's authentication key.
As a consequence, the authenticator and client users can all make sure 
each time whether the parties who share the entanglement pairs 
with themselves own the authentication keys or not and 
make sure that the secret key bits will
not be revealed or be inferred from the quantum or classical channels 
in our mutual authentication QDC protocol.

Besides having the ability to discover the possible different attacks from Eve, 
our authentication scheme can also 
avoid the attack by Trent if he prepares
different initial states and 
tries to steal the client users' messages. 
In the protocols by Lee \textit{et al.} \cite{lee:042305} and by Zhang
\textit{et al.} \cite{zhang:026301} as well 
as in our protocol, Trent, as an authenticator,
is considered to be more powerful than the rest of other parties or
users since all the users' secret identities are known to him, 
and all the quantum resources are issued by him. 
Thus in our protocol, we use a mutual authentication scheme in which a
user possesses extra ancilla qubits, can perform CNOT gates between his/her 
qubits and the ancilla qubit, and perform local operations ($I/H$ and $I/X$) and quantum measurements on the ancilla qubits when the user authenticates Trent. 
This authentication process may appear
slightly more complicated than that of the QKD-like scheme and 
than that of the protocols by Lee \textit{et al.} \cite{lee:042305}
and by Zhang \textit{et al.} \cite{zhang:026301}. 
But the way that the user can issue more quantum
resources (extra ancilla qubits and manipulations and measurements on the ancilla qubits) when authenticating Trent is the key point in our protocol
to prevent the attacks by the authenticator Trent if he prepares
different initial states, while the above mentioned protocols fail to prevent
(see, e.g., discussions in Sec.~\ref{Sec:TrentAttacks} and in Refs.~\cite{lee:042305,zhang:026301}).
If now suppose Trent prepares initial GHZ states 
$\frac{1}{\sqrt{2}}(\ket{00}+\ket{11})_{TEA}$
instead of EPR states that he is supposed to prepare. 
This unfaithful action of Trent is similar to the Eve's {\it intercept-and-CNOT attack} mentioned above, but the difference is
that now Trent knows the authentication keys. 
The QKD-like with an authenticator scheme will be 
vulnerable to this initial GHZ state attack by Trent 
(though the detailed steps of how this may happen are not shown here). 
We show below that this illegal action of Trent
will be discovered in the verifying set $V$
of the checking Step 6 of our
authentication process illustrated in Sec.~\ref{Sec:authentication}. 
The checking procedure starts from Alice's \textup{CNOT} operation on her 
particle $A$ and the prepared ancilla particle $a$. This operation will entangle
the three particles in the GHZ state with the ancilla particle, 
and will result in a state 
expressed as $\frac{1}{\sqrt{2}}(\ket{0000}+\ket{1111})_{TEAa}$. The
next step will depend on the bit value of the shared secret key
$AK_A$. 
When the $i$th bit value of $AK_{A}$ is $1$,
Alice will make $H$ operations separately 
on her two qubits (particles), i.e., $A$
and $a$, in the $i$th position in the verifying set $V$. 
For the purpose of discovering Trent's illegal action, 
there is no difference here whether Trent will follow the protocol to
make his subsequent quantum operations or not.
For simplicity, we suppose that Trent follows the protocol and  does
the same $H$ operations on his qubits 
when the $i$th bit value of $AK_{T}$ is $1$. 
Suppose now that the the $i$th bit value of $AK_{T}$ is $1$.
The state of the four qubits will become
$\frac{1}{\sqrt{2}}\{[(\ket{00}+\ket{11})\ket{0}+(\ket{01}+\ket{10})\ket{1}]\ket{0}-[(\ket{00}+\ket{11})\ket{1}+(\ket{01}+\ket{10})\ket{0}]\ket{1}\}_{TEAa}$.
Alice then measures the state of the ancilla particle in the $Z$-basis, 
and if the measurement result is $0$,
she will do nothing before her next
$Z$-basis measurement on particle $A$. Otherwise, she will make an $X$ operation
on her particle $A$ before the $Z$-basis measurement. Therefore, after Alice
measures her ancilla particle and 
performs the subsequent $I$ or $X$ operation, the
remaining three-particle state will become either
$\frac{1}{\sqrt{2}}[(\ket{00}+\ket{11})\ket{0}+(\ket{01}+\ket{10})\ket{1}]_{TEA}$
or
$\frac{1}{\sqrt{2}}[(\ket{00}+\ket{11})\ket{1}+(\ket{01}+\ket{10})\ket{0}]_{TEA}$
, corresponding to 
ancilla's measurement result that is either $0$ or $1$, respectively. 
It is not hard
to see, for each bit value of $AK_A$ to be $1$, Trent's measurement
result will, half the time, not match Alice's measurement result. 
The main reason is that the ancilla particle (qubit) is prepared by the
verifier, Alice. Trent can neither operate on the ancilla qubit nor know its
measurement result, so he cannot dominate in the authentication process.
Since Alice will discover the illegal action of Trent (similar to the
existence of Eve despite he knows the authentication key) if Trent indeed prepares different initial states
in the authentication process, she
will stop the subsequent communication process
and thus her secret message will not leak out.

In summary, it may appear that both our scheme and the QKD-like scheme 
require some shared authentication keys (secrets)
to begin with to perform mutual authentication, though they are
used in different ways. 
But one of significant differences is that the
secret key bits in the QKD-like scheme that uses the
Wegman-Carter authentication scheme~\cite{W-C:1981} need to be
different each time and thus eventually need to be refreshed
(replaced) in order to guarantee the 
absolute security of the authenticated classical channels \cite{bb84}.
If one uses quantum channels to refresh the shared secret
keys, this will make the QKD-like scheme be exactly 
similar to formal QKD rather than QDC that is intended here.
Our QDC protocol may, however, use the same shared key bits each time 
without compromising the system's security, at least in the 
possible attacks by Eve analyzed above.
In addition, we have pointed out that the QKD-like scheme with direct mutual authentication between any two users may not be practical in the implementation of a realistic quantum communication networks. So a scheme with an authenticator Trent who not only provides EPR pair qubits but also involves in the authentication process should be considered in the QKD-like scheme. 
Furthermore, the QKD-like with an authenticator scheme may still be 
vulnerable to the attack by the authenticator, if the authenticator prepares
different initial states (though the detailed steps of how this may happen are not shown here). 
Our QDC protocol, on the other hand, 
can discover this attack of Trent's illegal action 
and can prevent the secret message from leaking out.

\section{Conclusion}
\noindent
To summarize, it has been shown that the protocols proposed by Lee
\textit{et al.} \cite{lee:042305}
and the improved version by Zhang \textit{et al.} \cite{zhang:026301} cannot
prevent the authenticator Trent from knowing the secret message.  
To overcome these problems, we have presented a new quantum protocol
that uses the resources of the Bell states, the 
local operations and the entanglement swapping. 
In our proposed QDC 
protocol, the message communication process only starts
after the successful authentication process.
The authenticator Trent, after finishing his authentication job, will
leave the users alone to communicate with each other and to send the
secret message between themselves. 
Our protocol hence can prevent the real authenticator Trent from knowing the
secret message, a problem that the protocols proposed by Lee
\textit{et al.} and Zhang \textit{et al.} fail to resolve.  
The Bell measurements
by Trent in the communication process will cause the
entanglement swapping. The authenticated users/parties can then communicate
with each other securely with the resources of the entangled Bell pairs
between them. 
In the message transmission process, 
the concept of the local unitary operations and the entanglement
swapping is again used to encode and transmit the secret message. 
So no direct quantum link is required between
any two users, say Alice and Bob, who want to communicate with each
other. This might be an appealing advantage in the
practical implementation of a realistic quantum communication network.
It also avoids possible eavesdroppers to gaining any meaningful 
information of the secret message 
in the communication process.
The authenticator Trent can do almost everything
in an authentication network, 
the mutual authentication is therefore introduced 
in our protocol to prevent the attacks from an imposter Trent.
Our mutual authentication protocol can thus achieve 
secure QDC provided that the authenticator
Trent will do his authentication job faithfully.
The protocols proposed by Lee
\textit{et al.} and Zhang \textit{et al.}, on the other hand, 
also fail to prevent illegitimate
party to step in and act as the authenticator.
If, however, the genuine authenticator Trent would ask a fake
Bob to receive the secret message from 
Alice in our protocol, this could also be possibly prevented by allowing the
users/clients to access the classical public
channel at any time. If someone pretends to be Bob to communicate with Alice,
the real Bob may discover this event during the attack.

\nonumsection{Acknowledgments}
\noindent
H.S.G. would like to acknowledge support from the National Science
Council, Taiwan, under Grants No. 97-2112-M-002-012-MY3, 
support from the Excellent 
Research Projects of the National Taiwan University under 
Grants No. 97R0066-65 and No. 97R0066-67,
and support from the focus group
program of the National Center for Theoretical Sciences, Taiwan.
C.A.Y. and S.J.H. would like to acknowledge support from the National 
Science Council, Taiwan, under Grants No. 97-2221-E-239-022- and 95-2221-E- 
011-032-MY3.

\nonumsection{References}
\noindent

%
%
%

%

\end{document}